\documentclass[aps,pre,longbibliography,notitlepage,floatfix,nofootinbib,showkeys]{revtex4-1}

\usepackage{bm,amsmath,amssymb,graphicx,stmaryrd,float,subcaption,upgreek,sidecap,MnSymbol,mathtools,color}
\usepackage{epstopdf}
\usepackage[toc]{appendix}
\usepackage[justification=raggedright]{caption}
\usepackage[abs]{overpic}
\newcommand{\uvc}[1]{\bm{\mathrm{\hat #1}}} 
\usepackage{comment,footnote}
\usepackage{xcolor}

\newcommand{\bx}{{\bf x}}

\newcommand{\bn}{{\bf n}}
\newcommand{\bmo}{{\bf m}}
\newcommand{\bP}{{\bf P}}
\newcommand{\bJ}{{\bf J}}

\usepackage{hyperref}

\begin{document}
	\title{Exact and approximate mechanisms for pure bending of sheets}
	\author{Tian Yu}
	\email{tiany@princeton.edu}
	\affiliation{Engineering Mechanics Graduate Program,\\ Virginia Polytechnic Institute and State University, Blacksburg, VA 24061}
	\author{J. A. Hanna}
	\email{jhanna@unr.edu}
	\affiliation{Mechanical Engineering, University of Nevada, Reno, NV 89557\\}
	\date{\today}
\begin{abstract}
Direct measurement of the moment-curvature response of sheets or wires up to high curvatures can aid in modeling creasing, pleating, and other forming operations.
We consider theoretical issues related to the geometry of pure bending.
We present a linkage design that, for homogeneous deformation of the sample, results in an exact pure bending state up to arbitrarily high curvature, by combining a cochleoidal trajectory with an angle-doubling mechanism.
The full mechanism is single degree of freedom and position controlled, a desirable feature for measurement of soft materials.
We also present an optimal approximation to the cochleoid using a more easily implemented circular trajectory, and compare this with an existing commercial system for fabric testing.
We quantify the error of the approximate test by calculating the deformation of an Euler \emph{elastica}, a structure with linear moment-curvature response.
While the circular mechanism can approximate the exact boundary conditions quite well up to moderately large curvatures, the resulting curvature of the test sample still deviates significantly from homogeneous response.  
We also briefly discuss expectations of localization behavior in pure bending tests. 
\end{abstract}

\maketitle

\section{Introduction}\label{intro}

The response of sheet materials to bending is an important element in understanding diverse operations, such as the forming of metallic panels, and pleat and crease formation in fabrics or deployable structures made of composites or polymers.
While established standards exist to measure elastic bending moduli at low curvatures or through \emph{elastica} theory for larger deformations, the measurement of elastic-plastic moment-curvature response at high curvatures has been attempted through a variety of disparate approaches.

Ideally one would like to predict deformation in a forming process, involving complex loadings, by employing constitutive information obtained from one or a few simple mechanical tests.  A well-known example of such a test involves uniaxial tensile deformation, wherein the ends of a sample are displaced and the material experiences a uniform elongation, unless its properties are such as to promote strain localization.
In the present work we consider the analogous operation of pure bending, wherein both the position and orientation of the ends of the sample are prescribed so as to allow a uniform curvature, so long as material properties permit. We detail an exact and a related approximate mechanism for generating this deformation, in which a uniformly deforming sample adopts 
the shape of circular arcs of decreasing radius. 
The resulting moment-curvature curve \cite{Marciniak2002mechanics} can serve as the primary input to modeling of thin sheet forming processes where in-plane loading and stretching are minimal, such as some pleating and folding operations in which the sheet can be approximated as inextensible.
The current work does not address combined stretching and bending conditions which may be important for some applications.

Direct measurement of the moment-curvature response of a sheet is preferable over inference of the bending properties from tensile and compressive measurements for two reasons.  First, the sheet material may have complex anisotropic constitutive properties such that reconstruction of the bending behavior is difficult.  Second, it is not feasible to measure compressive response of a thin sheet, and for materials produced only in sheet form there may not exist any corresponding bulk object whose compressive response can be measured.
In fact, there are situations in which it might be more feasible to infer tensile and compressive behavior from bending measurements \cite{laws1981derivation, mayville1982uniaxial} than the other way around.

The mechanics of crease formation and recovery have application to textiles and folded deployable structures.
Approaches to this question in both fields are typically based on a test geometry in which a piece of material is simply compressed between rigid platens \cite{steele1957method, stuart1966investigation, stuart1967investigation, yee2005folding, SatouFuruya11, abbott2014characterization, dharmadasa2018characterizing, Elder19}, although other approaches to creating a crease have been employed \cite{bostwick1962comparison, skelton1971bending, Benusiglio12, Secheli19}.
Extracting moment-curvature data from these or other non-uniform bending tests requires either visual measurements of geometric features or an interpretation of the data in the light of some constitutive information \cite{clapp1990indirect, sanford2010large, dharmadasa2018characterizing, Sharma19, Secheli19}.

The earliest position controlled pure bending device of which we are aware is that of Isshi \cite{isshi1957bending}, which employed a cochleoidal groove and a set of gears to guide the position and orientation of one end of a fiber or fabric sample.  A variant of this device was constructed by Popper and Backer \cite{popper1968instrument}.
The Kato Tech KES-FB-2 machine performs a related approximate test as part of a suite of measurements known in the woven fabric community as the Kawabata Evaluation System (KES).  Though marketed as a ``Pure Bending Tester'', the instrument manual indicates a circular, rather than exact cochleoidal, path for the sample end, along with a set of gears and a compensating crank for orientation purposes \cite{KESmanual}.
In this paper, we present both a design for a linkage to perform an exact pure bending test, as well as an optimal circular approximation, either achievable with a single actuation.
In contrast to a uniaxial tensile test, the choice of whether position and orientation control are achieved by manipulating one or both ends of the sample has nontrivial effects on the geometry of the resulting motions.  One possible approach to pure bending is a synchronized rotation of both, and linear translation of one or both, ends of the sample using a minimum of three actuators \cite{koyama1990development}. 
A more complicated choreographed motion was implemented by Hoefnagels and co-workers in the context of \emph{in situ} SEM of thin films \cite{hoefnagels2015small}.

In theory, pure bending can also be achieved through force control, for example by applying a pure moment to one or both ends of a sheet sample while allowing the end to end distance to change freely.  Frictional resistance to this relative motion will prevent achievement of a pure bending state, so such constructions are not ideal for soft materials or weak structures.
An early construction by Eeg-Olofsson applied a moment to a textile sample through the action of a magnetic field on a coil floating in a pool of mercury  \cite{Olofsson1959some}.
Seffen and Pellegrino tested composite structures with a device that controls the orientation of the sample ends while allowing them to slide freely on a linear track \cite{seffen1999deployment}. 
An interesting device for composites testing is that of Murphey and co-workers, a fixture for a universal testing machine featuring rotating clamps, ball bearings, and rails \cite{sanford2011high, peterson2013large,murphey2015large}. 
There are other force controlled bending testers that rely on free movement of elements \cite{arnold2003pure, boers2010contactless}, sometimes aided by interconnected pulleys \cite{hill1972engineering, munoz2012simple, perduijn1995pure}.

Many approximate bending tests have been devised for sheet metals and other materials.  Some seek to extend the classic four-point test to moderate curvatures by allowing some freedom of motion at the sample ends \cite{kyriakides1987inelastic, zineb2003original, cimpoeru1993large}.
Fixtures for universal testing machines have been devised to achieve moderate or high curvature bending \cite{MarciniakKuczynski79, Duncan99-1, Duncan99-2, Weiss09, fernandez2018simple}.  The curvature distribution in such samples is more uniform than in simple compression between two platens, but is still far from uniform.
Other works \cite{Brunet01, Carbonniere09} that bend metals up to high curvatures provide little detail about the geometry. 
Other approaches include end loading with free rotation \cite{wisnom1992high}, and 
 direct position control \cite{goldacker2002bending, harris2008pure}.

In this paper, we develop an exact single-degree-of-freedom linkage design that can subject a sheet to pure bending, by augmenting a mechanism to generate motion along a cochleoid with an angle-doubling mechanism.
We also present a similar bending mechanism by replacing the cochleoidal motion by a simple circular approximation, and show that an existing commercial fabric tester is close to optimal among this class of approximations.
We compare exact and approximate approaches using the simple Euler \emph{elastica}, and reveal a surprisingly significant deviation in curvatures between the two.
We also briefly discuss localization in pure bending. 

\section{The geometry of pure bending, and an exact mechanism} \label{geometry}

\begin{figure}[htp]
	\centering
	\includegraphics[width=2.5in]{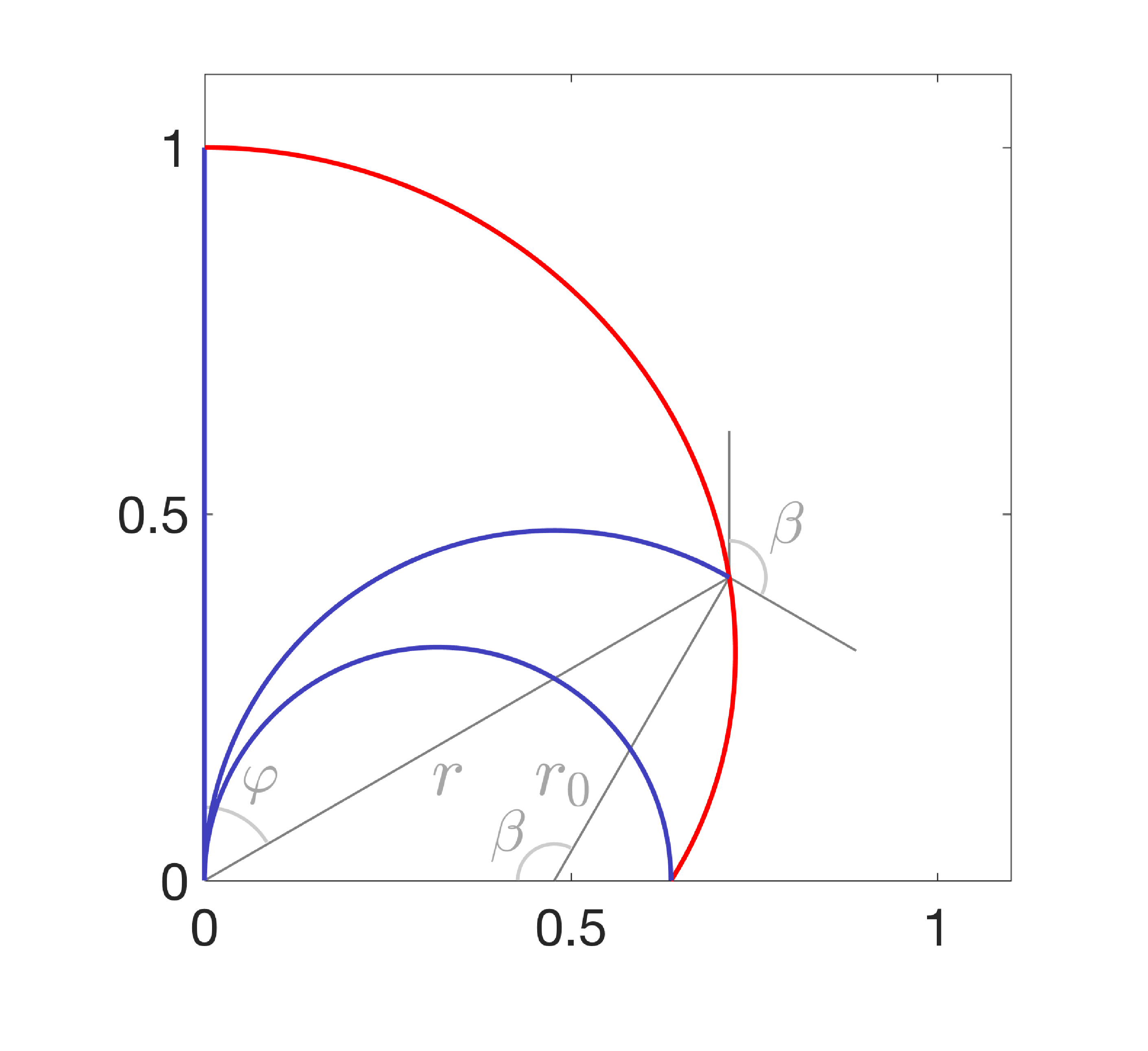}
\vspace{-0.1in}
	\caption{Part of a cochleoid (red curve) describing the locus of endpoints of an arc of length $L=1$ (blue curves) of progressively decreasing radius of curvature $r_0$.
	Also shown are the bending angle $\beta = L/r_0$ and the polar coordinates of the cochleoid $r$ and $\varphi$, with the latter measured clockwise from the vertical.}
	\label{cochleoidgeometry}
\end{figure}

A uniaxial tensile test imposes an increasing distance between the ends of a straight specimen.  
Under these conditions, some insufficiently work-hardening materials will experience localized deformation in the form of a necking instability.  However, most specimens will, for some range of strains, respond with an ideal uniform elongation, allowing measurement of the material's tensile stress-strain curve.
Analogously, a pure bending test can provide the moment-curvature response of a sheet if it can impose a deformation such that the sample can respond with an ideal uniform curvature.  This means that samples should adopt the shape of circular arcs of decreasing curvature.  If one end of the sample is fixed, the motion of the other end will be along a curve known as a cochleoid \cite{cochleoid2}, while the end orientations must also adjust accordingly.  The relevant geometric parameters are shown in Figure \ref{cochleoidgeometry}.
The polar equation of a cochleoid traced by a sample of length $L$ is 
\begin{equation}\label{polarequation}
r(\varphi)= L \frac{\sin \varphi}{\varphi} \, ,
\end{equation}
where $\varphi$ is measured clockwise from the fixed, vertically-oriented end of the sample.  The corresponding bending angle $\beta=2 \varphi$ is inversely related to the radius of curvature of the sample, $r_0 = L/\beta$.

We will first present a mechanism to trace the cochleoid, 
and then add another mechanism to double $\varphi$ to $\beta$ at the moving end.

\newpage

\subsection{A mechanism to generate the cochleoid} 
\label{mechanism}

\begin{figure}[h!]
	\centering
	\captionsetup[subfigure]{labelfont=normalfont,textfont=normalfont}
	\hspace{-1in}
	\begin{subfigure}[t]{0.48\textwidth}
		\centering
		\includegraphics[height=1.7in]{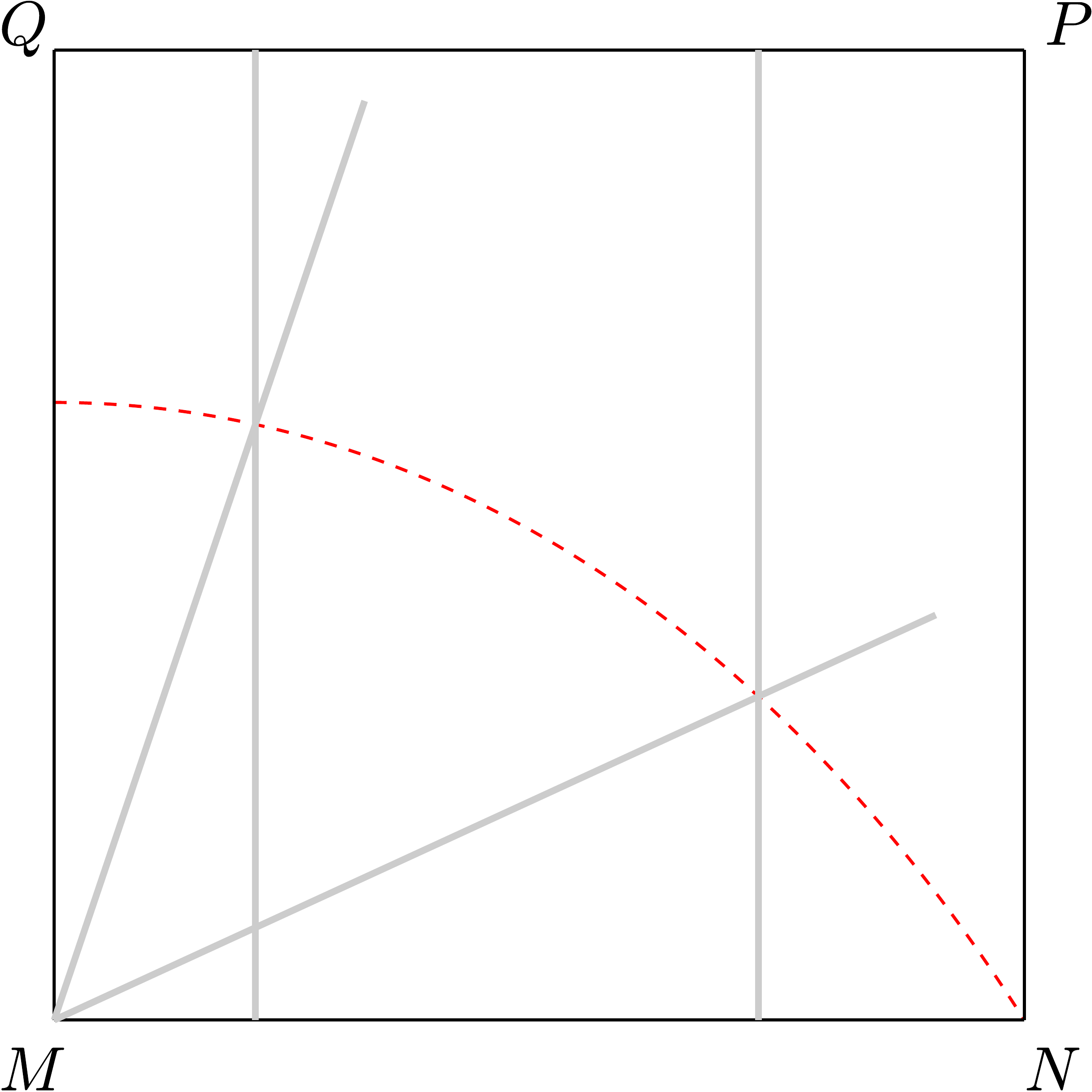}
		\caption{}\label{quacompass}
	\end{subfigure}
	\hspace{-0.75in}
	\begin{subfigure}[t]{0.48\textwidth}
		\centering
		\includegraphics[height=2.5in]{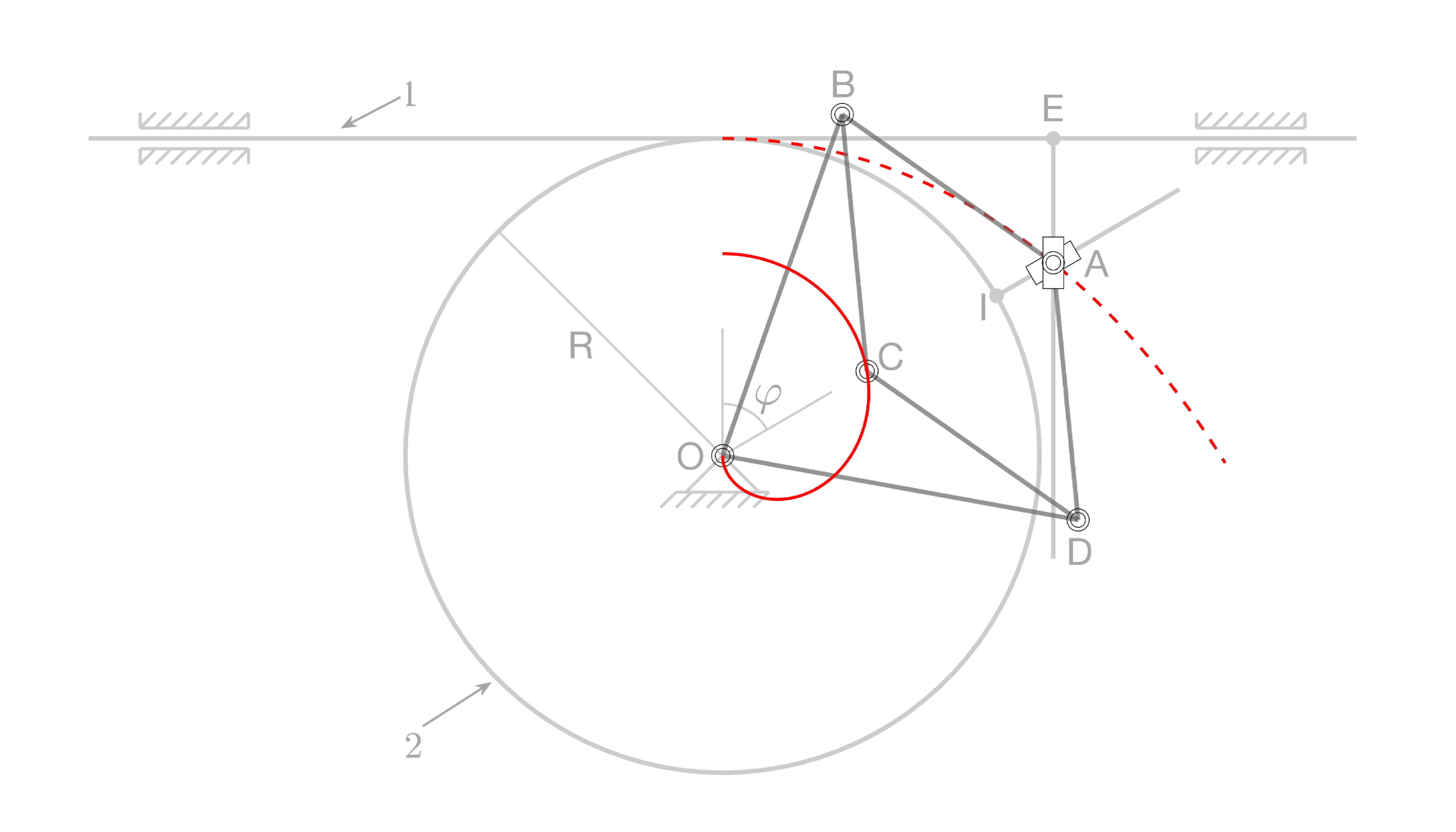}
		\caption{}\label{rotationmechanism}
	\end{subfigure}
	\caption{(\subref{quacompass}) The cochleoid's inverse (dashed red) is given by the locus of intersection of two lines, one initially along NP and translating, the other initially along MN and rotating about $M$, such that the two later coincide along QM.  (\subref{rotationmechanism}) After \cite{artobolevskii1964mechanisms}.  A rack and pinion system (light grey) generates the dashed red curve through translation of bar EA and rotation of bar IA.  This curve is inverted into the cochleoid (solid red) by a Peaucellier-type linkage OABCD (grey).
	}\label{cochleoidmechanism}
\end{figure}

The cochleoid sub-mechanism follows the example of Artobolevskii \cite{artobolevskii1964mechanisms}, which exploits the inversive relationship between the cochleoid and a special curve known either as the quadratrix of Dinostratus or the trisectrix of Hippias, names that reflect its other potential uses \cite{heath1921history}.
The quadratrix is easily generated.  Its equation $\rho(\varphi)$ is defined by $r(\varphi)\rho(\varphi) = R^2_{\text{inv}}$ for inversion of the cochleoid with respect to a circle of radius $R_{\text{inv}}$.
As shown in Figure \ref{quacompass}, if a line initially on one side of a square (NP) translates parallel to itself, and another initially perpendicular line (MN) rotates (about M), so that the two lines coincide at their final position on another side (QM) of the square, the quadratrix is the locus of intersection of these two steadily moving lines \cite{heath1921history}. 
Figure \ref{rotationmechanism} shows a rack and pinion system (light grey) to move point A along the quadratrix (red dashed line), using the motion of bars EA and IA coupled through sliders at A, with EA fixed perpendicularly to the rack and IA attached to, and rotating with, the pinion wheel. 
 To this system, a Peaucellier-type inversor (grey) is attached at the fixed origin O at the center of the wheel and at point A, so that point C moves along the cochleoid (red solid line).  
The sample (not shown) has one end fixed at O and the other at the moving point C. This mechanism has only one internal degree of freedom \cite{artobolevskii1964mechanisms}.

The inversor links satisfy $\text{OB}=\text{OD} = l$, $\text{AB} = \text{BC} = \text{CD} = \text{DA} = m$.  The  inverting circle relating C and A has its center at O and a radius given by $R^2_{\text{inv}} = l^2-m^2$.  Points O, C, and A are collinear.
As $r(\varphi)$ and $\rho(\varphi)$ respectively describe the positions of points C and A, a bit of trigonometry involving O, B, and the midpoint between A and C provides the expression
\begin{equation}\label{PauLip}
m^2 - \left(\frac{\rho (\varphi)-r(\varphi)}{2}\right)^2 + \left(r(\varphi)+\frac{\rho (\varphi)-r(\varphi)}{2}\right)^2=l^2 \, , \nonumber
\end{equation}
which upon simplification reveals that the wheel size $\rho(0)=R$ is related to the lengths of the links and the sample $r(0) = L$ by $LR=R^2_{\text{inv}} = l^2 - m^2$.

\subsection{Adding an angle-doubling mechanism}\label{fullbendingmechanism}

\begin{figure}[h!]
	\captionsetup[subfigure]{labelfont=normalfont,textfont=normalfont}
	\hspace{-1in}
	\begin{subfigure}[t]{0.48\textwidth}
		\centering
		\includegraphics[height=2.0in]{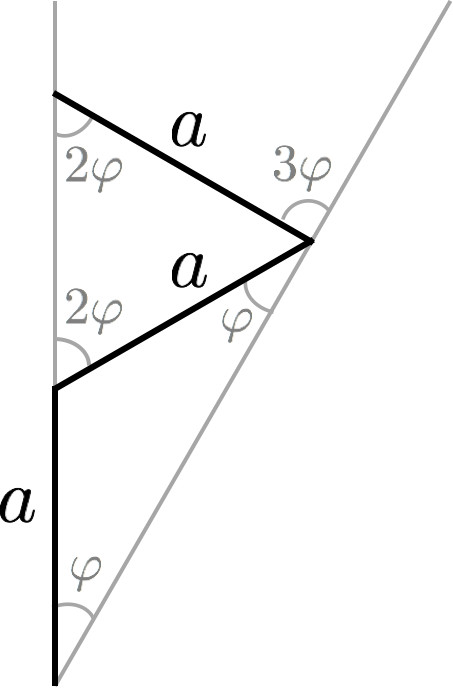}
		\caption{}\label{trigonometry}
	\end{subfigure}
\hspace{-0.75in}
	\begin{subfigure}[t]{0.48\textwidth}
		\centering
		\includegraphics[height=2.5in]{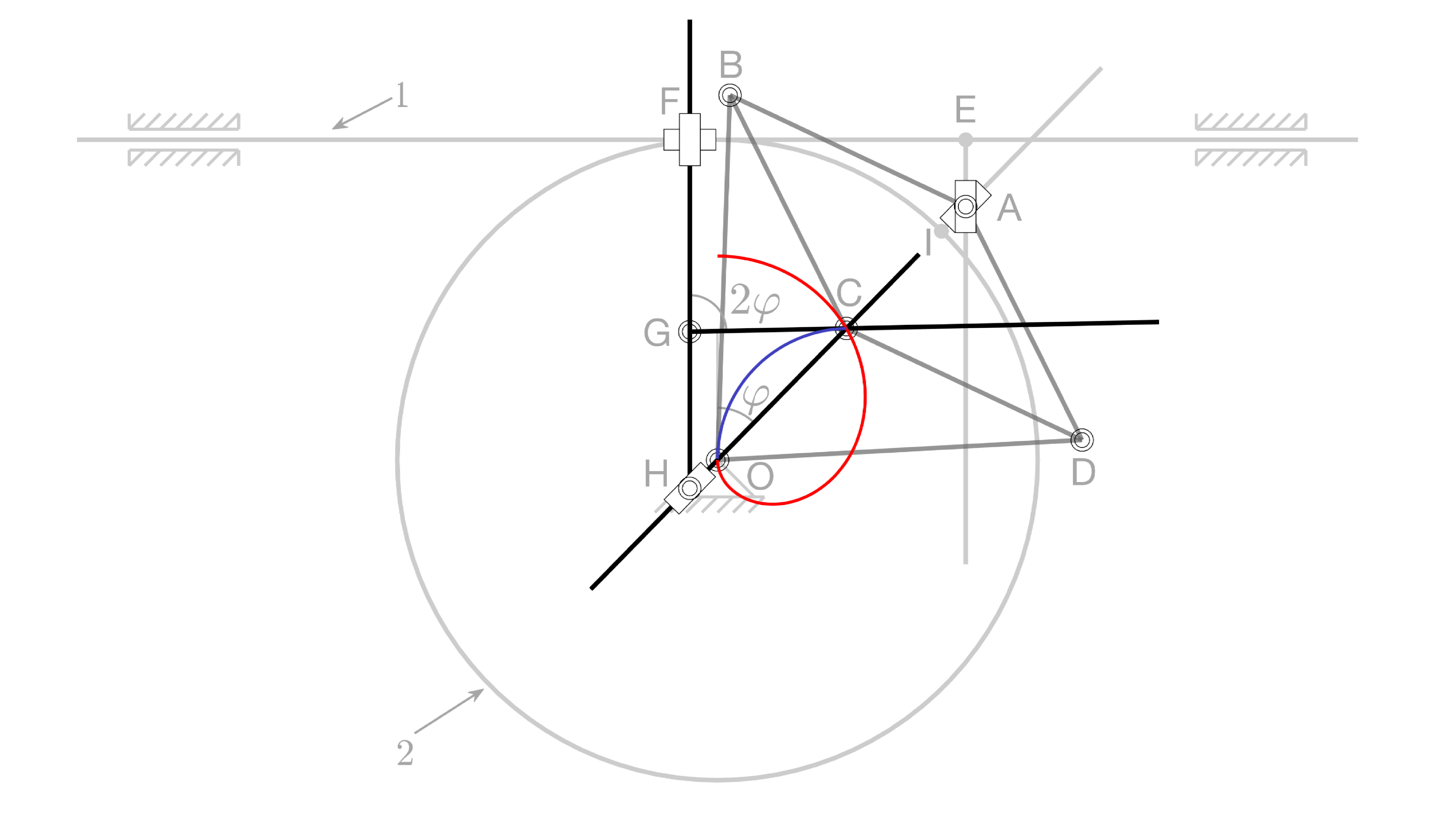}
		\caption{}\label{fullmechanism}
	\end{subfigure}
	\caption{(\subref{trigonometry}) Generating multiples of a fundamental angle $\varphi$ using equal length links in a wedge \cite{howround}.  (\subref{fullmechanism}) An angle-doubling mechanism (black) is added to the cochleoid mechanism of Figure \ref{rotationmechanism}.  Bar GC rotates to twice the angle of bar OC.	 The sample (blue arc) is clamped vertically at the fixed point O and in the orientation of bar GC at the moving point C.
	}\label{doubleanglemechanism}
\end{figure}

At the moving point C, the sample end must be oriented at twice the angle from the vertical turned by the wheel, $\beta=2\varphi$.
To double the angle $\varphi$, we exploit a simple construction \cite{howround} involving identical-length links arranged in a wedge as in Figure \ref{trigonometry}.  The lower triangle corresponds to CGH in Figure \ref{fullmechanism}, where an angle-doubling linkage (black) has been added to the cochleoid mechanism of Figure \ref{rotationmechanism}.  
Bar OC is attached to, and rotates with, the wheel, while bar GC rotates at twice its rate, so that it is oriented at $\beta = 2\varphi$ from the vertical.  The distances GH and GC are fixed and identical, bar GH is kept vertical through a double-slider connection at F on the rack, and the hinge-slider at H slides along bar OC. Note that point H passes over point C when $\varphi = \pi/2$.
The sample (blue arc) is clamped vertically at the fixed point O and in the orientation of bar GC at the moving point C.

The angular scope of the combined mechanism is limited by a singular parallel configuration adopted by the kite-shaped inversor at a finite angle.  The maximum curvature achievable depends on the choice of link lengths and wheel radius, and can be made arbitrarily large.  Details are shown in Appendix \ref{limitingconfigurations}.

The combined mechanism may be implemented as a multi-level structure. 
 It may also be practical to displace some elements laterally with respect to their positions in Figure \ref{fullmechanism}, namely the connection E and slider F which could sit on a separate parallel track instead of directly on the rack and pinion system.
Torque can be measured by a sensor at the origin O.
The mechanism has one internal degree of freedom, and can be driven either by a motor to rotate the 
gear or by a linear actuator to translate the rack.

\section{An approximate circular mechanism, and comparison with other testers}\label{approximategeometry}

The exact mechanism introduced in Section \ref{geometry} is complex.  It is worth investigating whether a simpler approximate mechanism might serve as a reasonable substitute.
The approximation we consider involves replacing the cochleoid with a circular arc, similarly to what is done in the commercial KES fabric tester \cite{KESmanual}, while retaining orientation control through an angle-doubling linkage as in Section \ref{fullbendingmechanism}.
The latter should be comparable or more accurate than the unspecified approximate angular compensation performed by the KES tester \cite{KESmanual}, and could in theory be replaced by a gear train as in Isshi's tester \cite{isshi1957bending}.

\begin{figure}[h!]
	\captionsetup[subfigure]{labelfont=normalfont,textfont=normalfont}
	{\hspace{-0.75in}
	\begin{subfigure}[t]{0.4\textwidth}
		\includegraphics[height=1.9in]{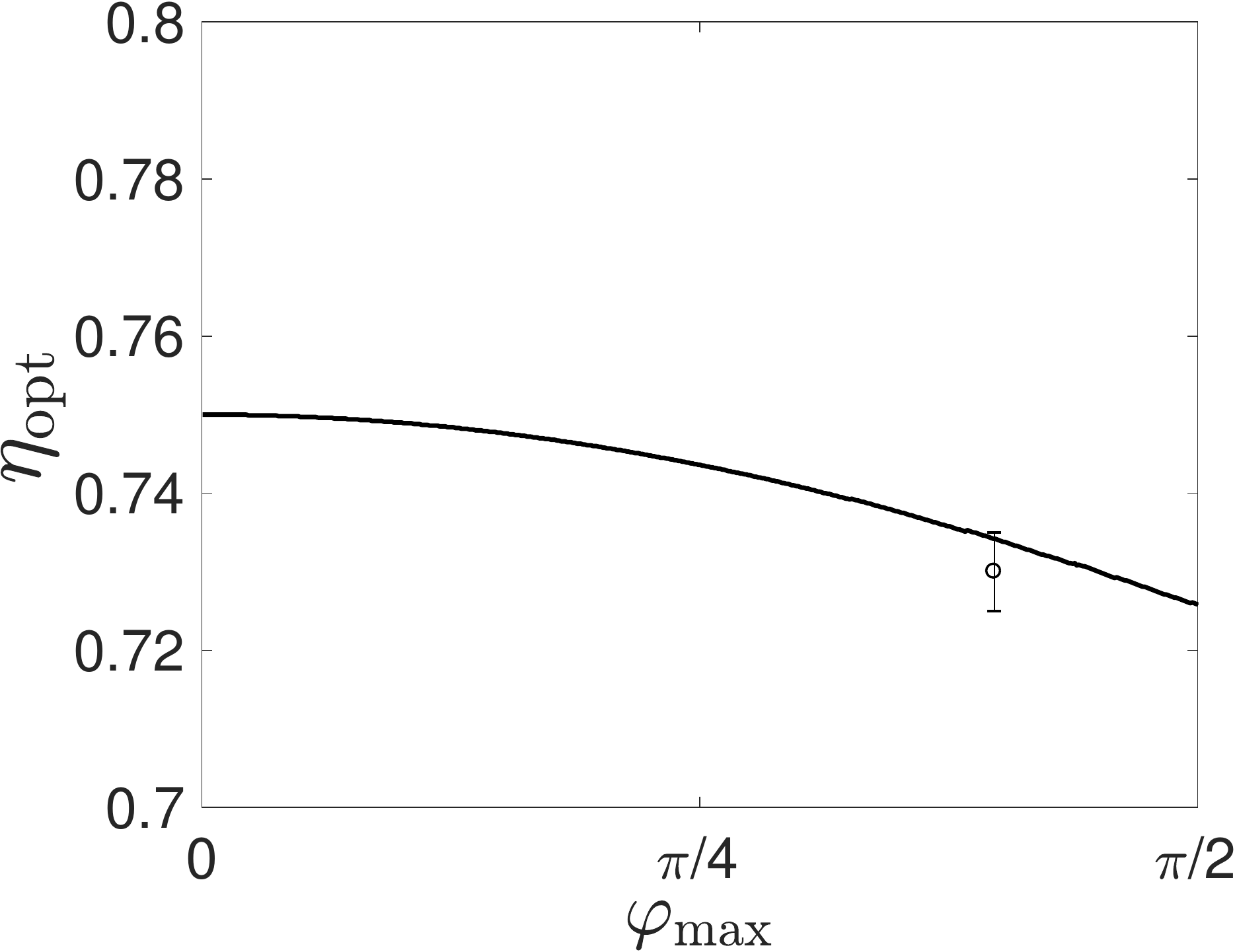}
		\caption{}\label{BestcircleVSphimax}
	\end{subfigure}
	\hspace{-0.3in}
	\begin{subfigure}[t]{0.35\textwidth}
		\includegraphics[height=1.9in]{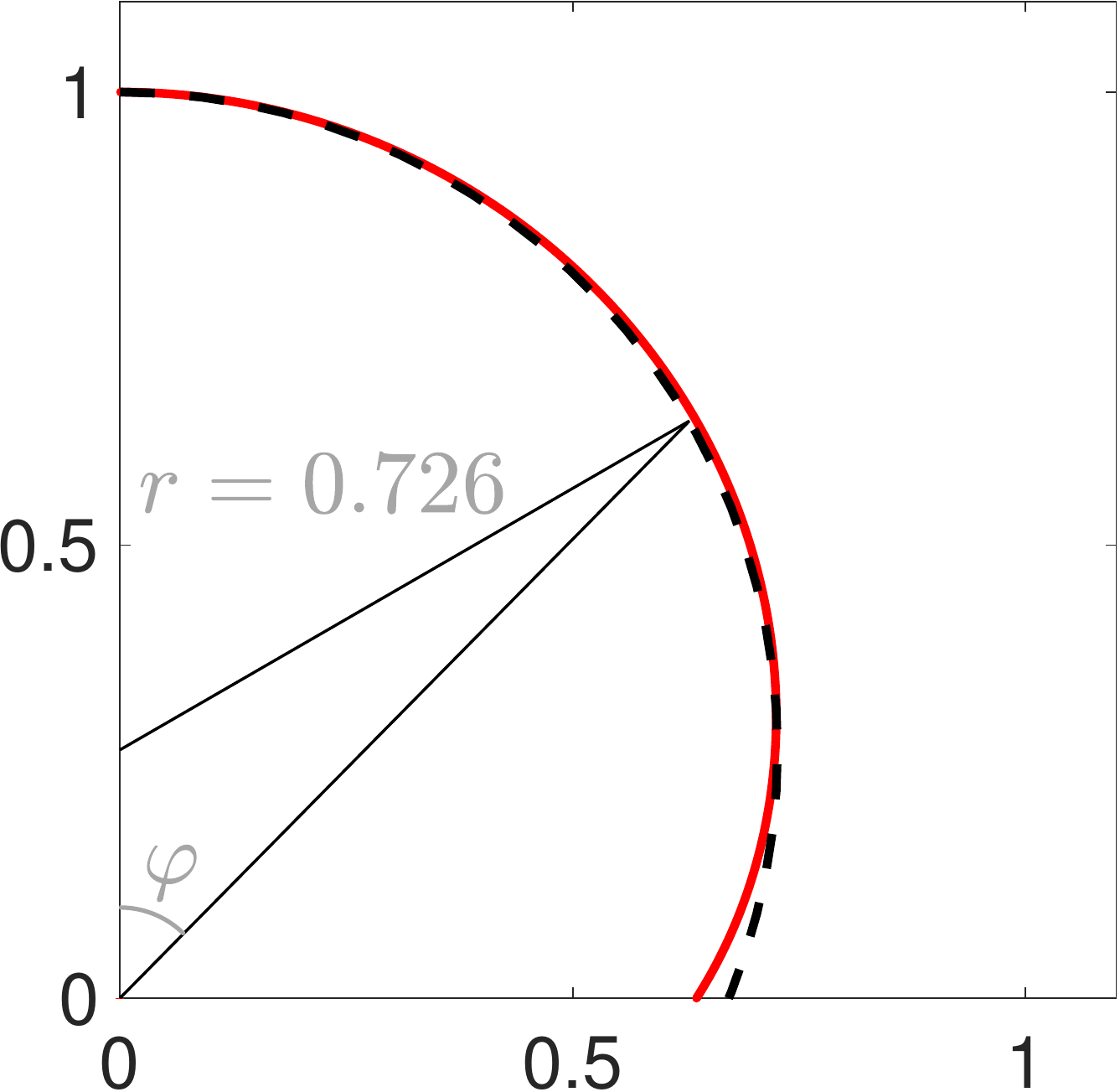}
		\caption{}\label{approxcircles}
	\end{subfigure}
	\begin{subfigure}[t]{0.2\textwidth}
		\includegraphics[height=1.9in]{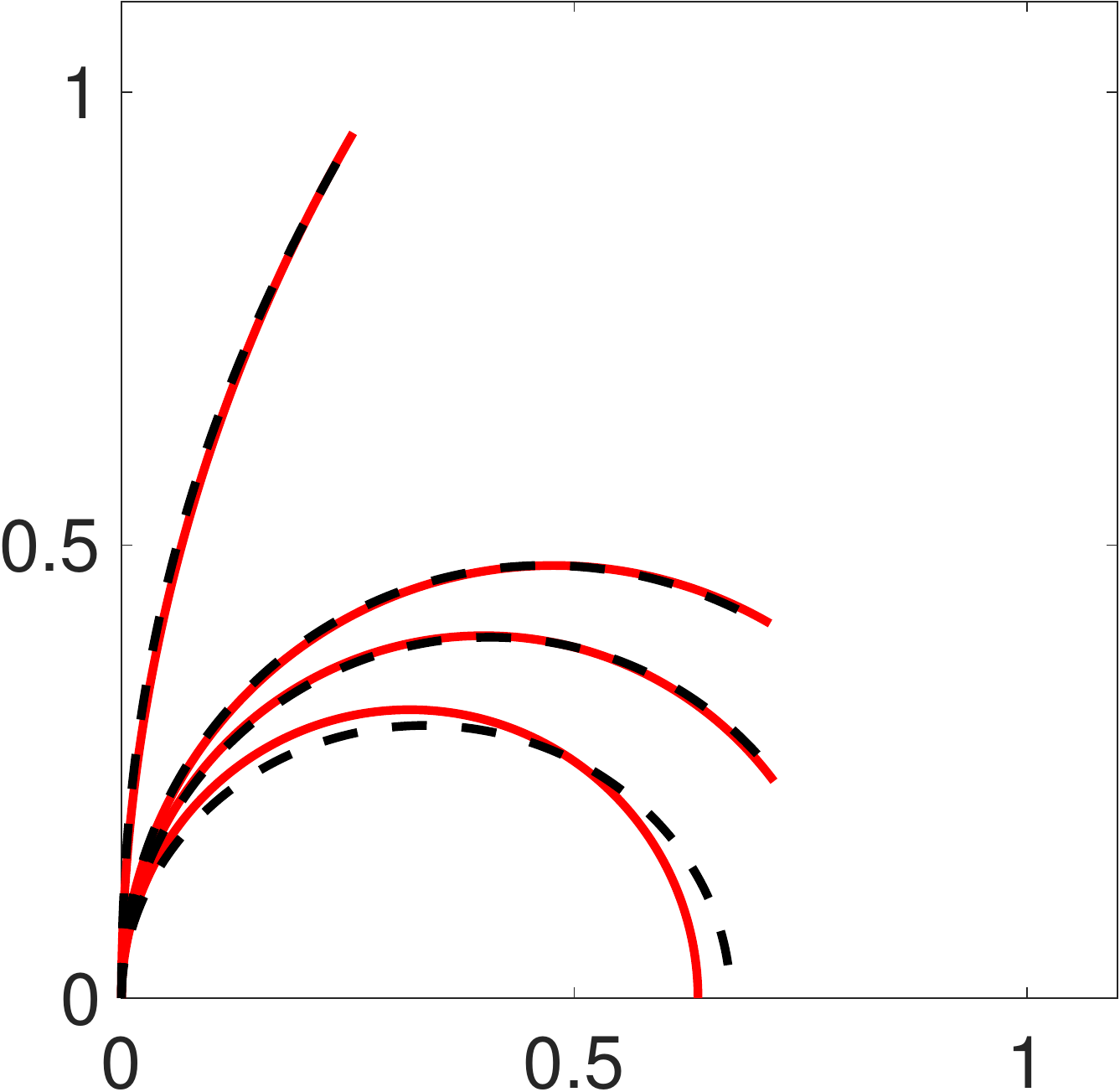}
		\caption{}\label{kesp726config}
	\end{subfigure}}\\
\vspace{0.1in}
	{\begin{subfigure}[t]{0.45\textwidth}
		\includegraphics[height=2.2in]{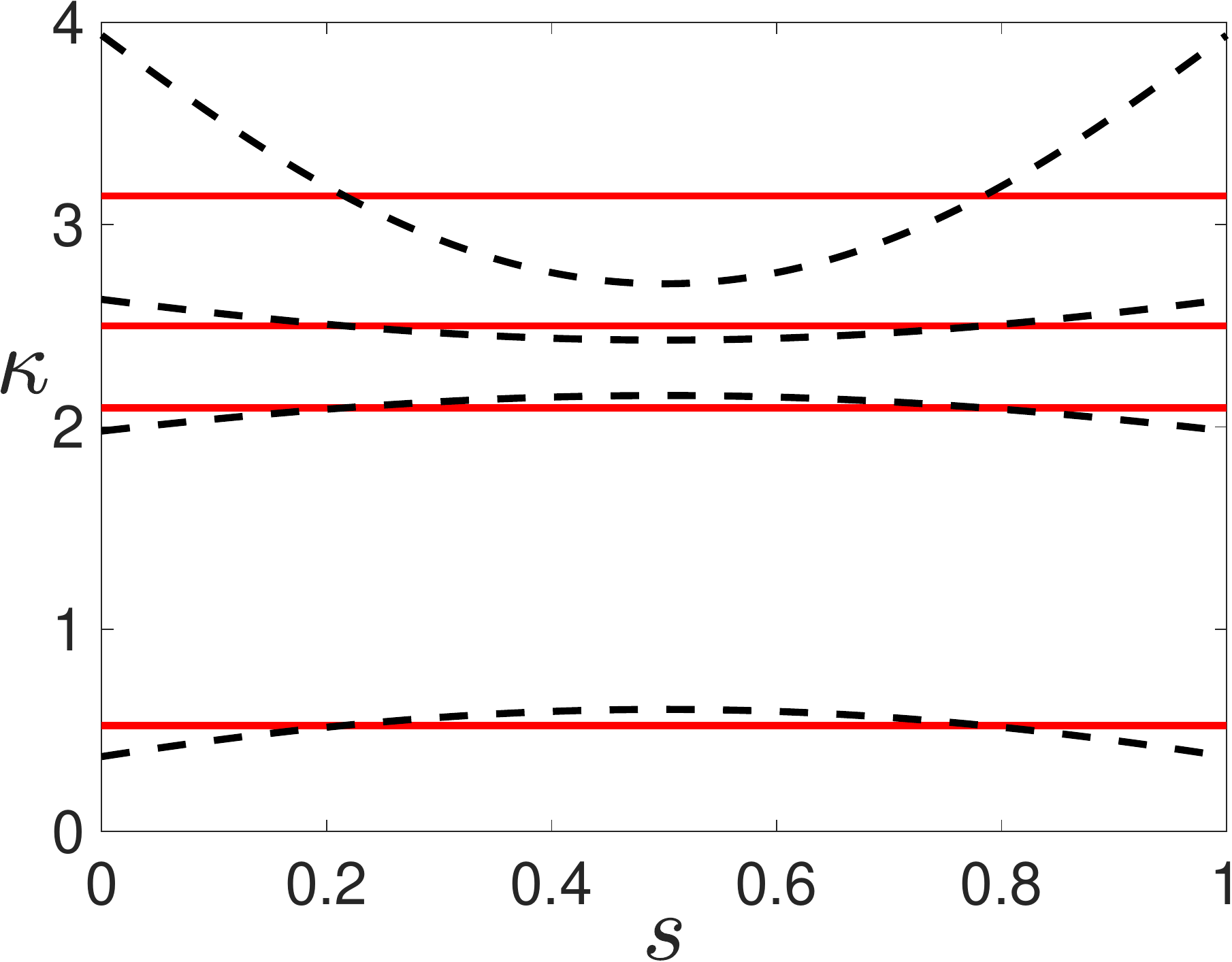}
		\caption{}\label{kesp726curvature}
	\end{subfigure}
	\begin{subfigure}[t]{0.45\textwidth}
	\hspace{-.3in}
		\includegraphics[height=2.2in]{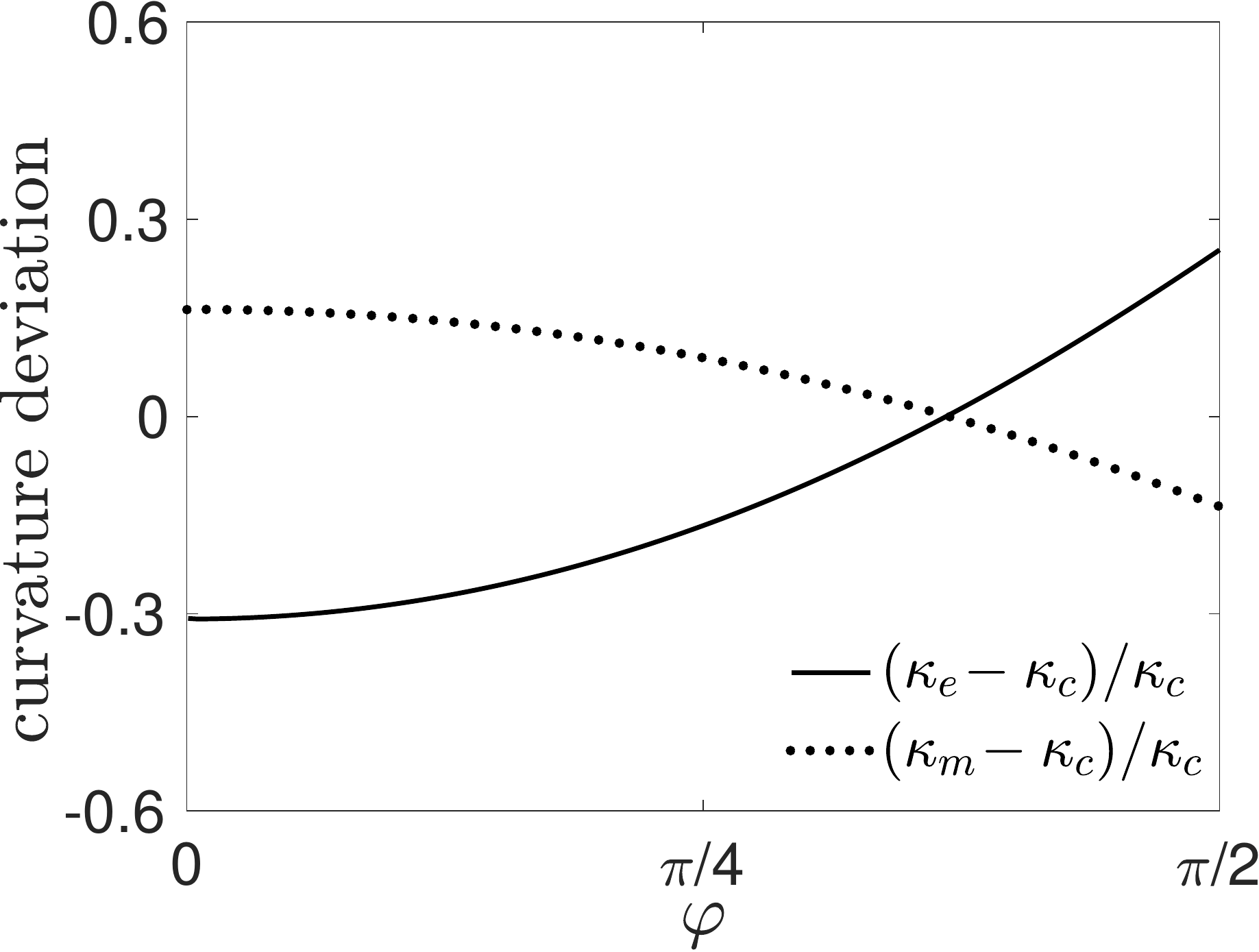}
		\caption{}\label{Curvdeviation}
	\end{subfigure}}
	\caption{ (\subref{BestcircleVSphimax}) Optimal radius $\eta_{\text{opt}}$ of the approximate circular trajectory as a function of the maximum bending angle $
\varphi_{\text{max}}$ reached by the tester.  The error is comparable to the thickness of the curve.
The single data point corresponds to the the KES value $0.73 \pm 0.005$ at $\varphi_{\text{max}}=\frac{5}{4}$.  Other subfigures use the value $\eta_{\text{opt}} = 0.726$ corresponding to $\varphi_{max} = \frac{\pi}{2}$ to compare the response of an Euler \emph{elastica} under exact (solid red) and approximate (dashed black) boundary conditions, including a mechanism to double $\varphi$.  (\subref{approxcircles}) Sample end positions. (\subref{kesp726config}) Sample configurations and (\subref{kesp726curvature}) Curvature distributions 
at bending angles $\varphi = \frac{\pi}{12}, \frac{\pi}{3}, \frac{5}{4}, \frac{\pi}{2}$.  
 (\subref{Curvdeviation}) Normalized curvature deviations at the ends (solid black) and middle (dotted black) of the sample as a function of the bending angle $\varphi$.	}\label{numericalexp}
\end{figure}

We restrict our consideration to circular trajectories that coincide with the cochleoid in the unbent state, and are symmetric with respect to bending in either direction.  This leaves only the circle's radius $\eta$ as a parameter; 
the polar equation for the circle is 
$r_a (\varphi ; \eta)=\sqrt{\eta ^2-(1-\eta)^2 \sin^2 \varphi} +(1-\eta)\cos \varphi$.  The circle's center does not sit at the origin, and the angle to be doubled is $\varphi$, which is not the angle turned by the circle.
 We designate the best approximation, for a particular maximum angle $\varphi_{\text{max}}$ reached by the tester, as that which minimizes the area between the approximate circle $r_a (\varphi ; \eta)$ and the ideal cochleoid $r(\varphi)$, 
$\int_{0}^{\varphi_{\text{max}}}\! \frac{1}{2} |\, r 
(\varphi)^2 - r_a (\varphi ; \eta)^2 | \, d\varphi$.  
Numerically determined values of the optimal radius $\eta_{\text{opt}}(\varphi_{max})$ are shown in Figure \ref{BestcircleVSphimax}, along with the KES value corresponding to $\eta\left(\frac{5}{4}\right) = 0.73 \pm 0.005$ (we note that the Kato Tech manual \cite{KESmanual} states a range of $\pm 0.05$, but we believe this must be a misprint).  The KES value is consistent with the optimal circle. 
For further analysis, we choose a larger maximum angle than that of the KES machine, corresponding to a half-circle configuration $\beta = \pi$, $\varphi_{max} = \frac{\pi}{2}$, whose optimal approximating circle has radius $\eta_{\text{opt}}\left(\tfrac{\pi}{2}\right) = 0.7257 \pm 0.0001$,  
which we truncate as $\eta = 0.726$.

Figure \ref{approxcircles} compares this approximate (dashed black) circular trajectory with the exact (solid red) cochleoidal trajectory.
The curves are quite close up to an angle of about $\varphi=\frac{5}{4}$, which is the maximum angle used by the KES tester \cite{KESmanual}.  
However, this is somewhat misleading.  
If we combine the approximate trajectory with a mechanism to double the angle $\varphi$, we can quantitatively compare the response of a particular material to the exact and approximate boundary conditions imposed.
We take as the simplest possible example the Euler \emph{elastica}, an inextensible planar curve with linear moment-curvature response.  The method of solution is detailed in Appendix \ref{numerics}. 
Figures \ref{kesp726config} and \ref{kesp726curvature} show configurations and curvature distributions $\kappa(s)$, where $s$ is the arc length along the curve, 
for bending angles $\varphi = \frac{\pi}{12}, \frac{\pi}{3}, \frac{5}{4}, \frac{\pi}{2}\,$; recall that the sample is bent to subtend an angle of $2\varphi$.  
From Figure \ref{kesp726curvature}, we see that the curvature error for an \emph{elastica} has a simple unimodal form, with roughly equal excess and deficit in curvature, and with an inversion at moderate curvatures such that the maximum of curvature changes from being at the middle of the sample to being at the edges as the imposed curvature increases.  
Figure \ref{Curvdeviation} shows the normalized end and middle curvature deviations $(\kappa_e - \kappa_c)/ \kappa_c$ and $(\kappa_m - \kappa_c)/ \kappa_c$, where $\kappa_e$, $\kappa_m$ and $\kappa_c$ are the end, middle, and ideal curvatures, for bending up to a half-circle configuration.  Despite very small differences in boundary positions up to $\varphi = \frac{5}{4}$, the sample curvature under the approximate mechanism can show significant deviation from the ideal value at low bending angles, on the order of 30\%, leading to large errors in measured moment-curvature response.  Thus, one would expect discrepancies between bending moduli inferred from such a test and standard low-curvature tests.

\section{Discussion}\label{discussion}

Exact information about the end orientations in the KES tester is not available, but we believe the resulting theoretical errors are at least as large as those of the approximate mechanism shown in Figure \ref{numericalexp}, which has a slightly better position approximation, leading to a better orientation approximation.

Aside from theoretical errors, experimental error can arise from difficulties in implementing any mechanism.  Given the sensitivity of the curvature distribution to small errors in end position and orientation, it is not surprising to see significant errors in bending strains even in testers with multiple synchronized actuators to control boundary conditions, as for example in Figure 9 of \cite{koyama1990development}.
Our first attempt at implementing the exact mechanism of Section \ref{geometry} has suffered similar issues due to excess play between components.

As the mechanism is intended for use on very soft sheets, we have treated the link elements as rigid lines in our analysis.  However, for testing of stiff materials, it is expected that machine stiffness effects will be considerably more complex than in a simple tensile test.  In either case, sufficiently thick links should be employed.
We have also implicitly assumed quasi-static motions of the mechanism, ignoring any inertial effects.

Our choice of linkage is not unique.  An alternate inversor, such as that of Hart, could be used to generate the cochleoid from the quadratrix.  
We could have employed two cochleoids, or approximate circles, to rotate the sample ends about the center. 

The geometry of a pure bend test raises questions not present in the uniaxial tensile test.  For example, if we wish to avoid sample failure near the grips, it is not clear what the equivalent of a ``dog bone'' shaped sample would be.  Reducing the cross section, and thus the bending resistance, would lead to a different local curvature that in turn affects the global geometry, and thus the local contact moment along the sample.  Another obvious question without an obvious answer is how to interpret localization.  What is the equivalent of the Consid{\`{e}}re criterion for pure bending?  The balances of force and moment on an inextensible elastic rod are given by the Kirchhoff equations, which in the absence of body or other distributed forces can be integrated with respect to arc length $s$ to obtain
\begin{align}
	\bn &= \bP \, , \label{forcebalance} \\
	\bmo(s) + \bx(s) \times \bn &= \bJ \, , \label{torquebalance}
\end{align}
where $\bx(s)$ is the position of the body, $\bn$ and $\bmo(s)$ are the local contact force and moment, and the constant vectors $\bP$ and $\bJ$ are the conserved force and torque \cite{SinghHanna19}.  As the body is only loaded at its ends, the force balance \eqref{forcebalance} indicates that the contact force $\bn$ is a constant.  
From the perspective of quasistatic flow localization \cite{SemiatinJonas84}, we consider perturbations of a base state parameterized by a small increment $\chi$, such that all quantities $() = ()_0 + \frac{d()}{d\chi}d\chi$.  Additionally noting that the contact moment is simply a function of the curvature $\kappa(s)$, we obtain a localization criterion from the torque balance
\begin{align}
	\frac{\partial\bmo(\kappa)}{\partial\kappa}\frac{d\kappa(s)}{d\chi} + \frac{d\bx(s)}{d\chi} \times \bP_0 + \bx_0(s) \times \frac{d\bP}{d\chi} = \frac{d\bJ}{d\chi} = 0  \, .\label{localization}
\end{align}
Whether this criterion is satisfied clearly depends on the loading process.
If the base state is pure bending, $\bP_0 = 0$, we measure $\bmo_0 = \bJ_0$ at the sample end, and $\bx_0(s)$ describes an arc of a circle with curvature $\kappa_0$, simplifying the equation \eqref{localization} to 
\begin{align}
	\frac{\partial\bmo(\kappa)}{\partial\kappa} = -\frac{\uvc{r}(s)}{\kappa_0} \times \frac{d\bP}{d\chi} {\bigg{/}} \frac{d\kappa(s)}{d\chi} \, , \label{localizationpurebend}
\end{align}
where the unit vector $\uvc{r}$ corresponds to a cylindrical coordinate system sharing its origin with the circular arc of the sample's base state.
Choosing a particular shape perturbation $\frac{d\bx}{d\chi}$, $\frac{d\kappa}{d\chi}$ follows immediately, as $\kappa^2 = \frac{d^2\bx}{ds^2}\cdot\frac{d^2\bx}{ds^2}$.  However, the induced contact force $\frac{d\bP}{d\chi}$ must be determined from a full solution of the Kirchhoff equations and boundary conditions for the system.  
Once we know this, we can in principle determine when localization occurs in terms of the slope of the constitutive curve $\frac{\partial\bmo(\kappa)}{\partial\kappa}$.
For a symmetric perturbation, we expect the constant vector $\frac{d\bP}{d\chi}$ to be aligned with the end to end vector, and $\frac{d\kappa(s)}{d\chi}$ to have either a maximum in the center or two maxima on the sides, something easily seen by playing with an elastic band.
To understand localization, we need to determine how each of these quantities scale with curvature.
For a simple Euler \emph{elastica}, perturbative analysis of the shape equation \cite{SinghHanna19} near a circular solution seems to imply that $\frac{d | \bP |}{d\kappa} \sim \kappa_0$, but we reserve further analysis for future work.

There appear to be few studies focused on the onset of plastic localization in forming operations where bending is the predominant deformation mode.
Triantafyllidis and Samanta considered localization in combined bending and stretching of axisymmetric sheets \cite{TriantafyllidisSamanta86}. 
Kyriakides and co-workers \cite{Kyriakides08} observed localization arrest and propagation in shallow bending of tubes made from a L{\"{u}}ders-banding steel, and observed initiation of localization near the sides, rather than in the center.

\section{Conclusions}

We have presented exact and approximate single-degree-of-freedom mechanisms for the imposition of pure bending.  
Curvature in an approximately bent sample can deviate significantly from the nominal constant value.
The test geometry raises interesting issues, particularly with regard to localization.


\section{Acknowledgments} \label{Acknowledgments}
This work was supported by The NonWovens Institute (project 16-197).  We thank N. A. Corbin, D. Link, and W. D. Hartley II for help building a prototype tester, H. Singh for helpful discussions, and S. Guevel and E. DenHartog for information about the Kato Tech KES-FB-2 machine.

\appendix

\section{Maximum achievable curvature}\label{limitingconfigurations}

\begin{figure}[htp]
	\centering
	\includegraphics[height=2.5in]{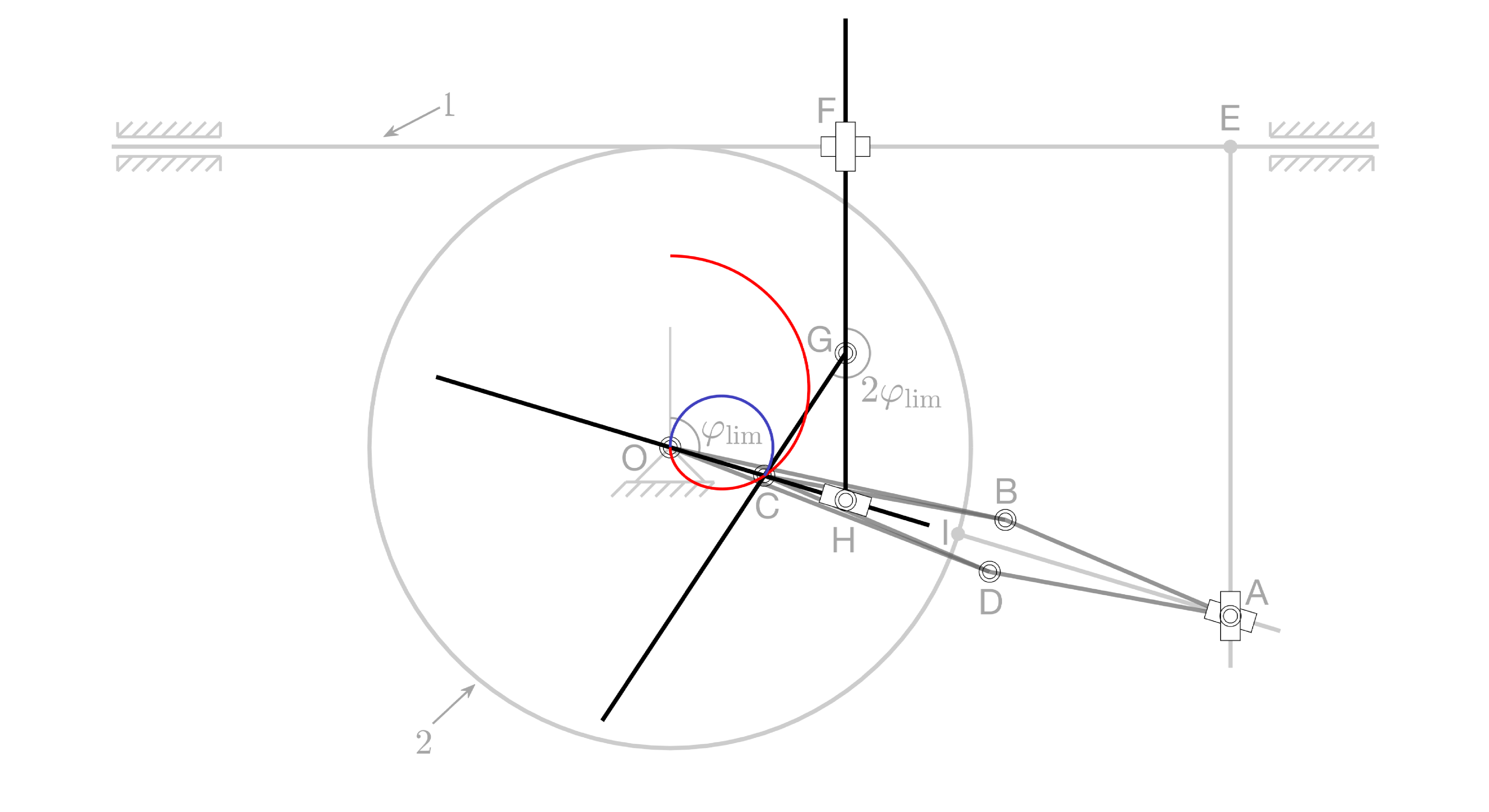}
	\caption{A configuration near the limiting configuration in which the links of the kite-shaped inversor OABCD align.}
	\label{limitconfig}
\end{figure}

The maximum curvature and minimum sample length are limited in practice by the bending angle at which the links of the kite-shaped inversor OABCD align (see Figure \ref{limitconfig}). This limiting angle $\varphi_\text{lim}$ is related to the link lengths $l$ and $m$ and the wheel size $R$ through the trigonometric relationship
\begin{equation}\label{limitequa}
\left(l + m\right)\sin\varphi_\text{lim} = R\varphi_\text{lim} \, .
\end{equation}
Note that $l+m < R$.  The sample radius at this limit is
\begin{equation}\label{limitcurvature}
r_{0\text{lim}}=\frac{L}{2\varphi_\text{lim}} =\frac{l^2 - m^2}{2R \varphi_\text{lim}} = \frac{l - m}{2 \sin \varphi_\text{lim}} \, ,
\end{equation}
where the second and third equalities in \eqref{limitcurvature} use $LR = l^2 - m^2$ from Section \ref{mechanism}, and \eqref{limitequa}, respectively.

Among other things, these relationships tell us that for a fixed $R$, the limiting radius can be made arbitrarily small at some finite limiting angle by shrinking the difference in link lengths $l-m$.

\section{Solutions for Euler $\emph{elastica}$}
\label{numerics}

The Euler \emph{elastica} is an analytically tractable system, but for expedience we decided to forego an extended exercise in elliptic integrals and instead use a continuation code for rods already on hand to quickly compute the solutions in Figures \ref{kesp726config}-\ref{Curvdeviation}.  The Kirchhoff equations for an inextensible rod in two dimensions, and the kinematics in terms of an angle $\theta$ measured clockwise from the vertical, are
\begin{equation}\label{Equilibrium}
\begin{aligned}
\partial_s N_2 -N_3 \kappa &=0 \, , \\
\partial_s N_3 +N_2 \kappa &=0 \, , \\
\partial_s M(\kappa) - N_2 &=0 \, , \\
\partial_s \theta &=\kappa \, , \\
\partial_s x &=\sin \theta \, , \\
\partial_s y &=\cos \theta \, , \\
\end{aligned}
\end{equation}
where $N_2$ and $N_3$ are the components of the contact force along the normal and tangential directions, respectively, and 
$M(\kappa)$ is a general constitutive law representing the moment-curvature response of the sheet.  For an \emph{elastica}, $M$ is linear in $\kappa$.  The choice of bending stiffness does not affect the shape in response to displacement boundary conditions.  
These conditions for the approximate mechanism are
\begin{equation}\label{Boundary2}
\begin{aligned}
x(0)=0,\, y(0)=0,\, \theta(0)=0,\, x(1)=\left(\sqrt{\eta^2-(1-\eta)^2 \sin \varphi ^2} +(1-\eta)\cos \varphi\right) \sin \varphi,\, \\
y(1)=\left(\sqrt{\eta^2-(1-\eta)^2 \sin \varphi ^2} +(1-\eta)\cos \varphi\right) \cos \varphi,\, \theta(1)=2 \varphi \, .\\
\end{aligned}
\end{equation}

We solve the system \eqref{Equilibrium}-\eqref{Boundary2} using the continuation package AUTO 07P \cite{doedel2007auto2}.  Starting with a half-circle configuration, we move the sample end to the position corresponding to $\varphi=\pi /2$, then continue $\varphi$ down to $0$, the straight configuration.  Using instead the straight configuration as a starting solution does not work well, perhaps because the equations are one-sidedly constrained by inextensibility.

\bibliographystyle{unsrt}

\end{document}